\documentclass[10pt]{article}

\usepackage{amsmath}
\usepackage{amssymb}
\usepackage{graphics}
\usepackage{rotating}
\usepackage{cite}
\usepackage{color}
\usepackage{fancybox}
\usepackage{pstricks}
\usepackage{xcolor}
\usepackage{shadowtext}
\usepackage{multicol}

\usepackage{multirow}

\usepackage{subfigure}
\usepackage{makecell}
\usepackage{diagbox}
\usepackage{colortbl}
\usepackage{hhline}

\usepackage{caption}

\usepackage[T1]{fontenc}

\def\0{\mbox{\tiny $0$}}
\def\1{\mbox{\tiny $1$}}
\def\2{\mbox{\tiny $2$}}
\def\3{\mbox{\tiny $3$}}
\def\4{\mbox{\tiny $4$}}
\def\5{\mbox{\tiny $5$}}
\def\6{\mbox{\tiny $6$}}
\def\7{\mbox{\tiny $7$}}
\def\8{\mbox{\tiny $8$}}
\def\9{\mbox{\tiny $9$}}
\definecolor{navy}{rgb}{0,0,.6}
\definecolor{jour}{rgb}{0,0.6,.4}
\definecolor{jbul}{rgb}{0.7,0.,.4}
\begin{document}
%
\thispagestyle{empty}
\setcounter{page}{0}

\begin{center}
\shadowrgb{0.8,0.8,1}
\shadowoffset{4pt}
\shadowtext{
\color{navy}
\fontsize{18}{18}\selectfont
\bf  MAXIMAL BREAKING OF SYMMETRY AT}\\
\shadowtext{
\color{navy}
\fontsize{18}{18}\selectfont
\bf CRITICAL ANGLE AND CLOSED FORM}\\
\shadowtext{
\color{navy}
\fontsize{18}{18}\selectfont
\bf EXPRESSION FOR ANGULAR DEVIATIONS}\\
\shadowtext{
\color{navy}
\fontsize{18}{18}\selectfont
\bf OF THE SNELL LAW}
\end{center}

\vspace*{1cm}

\begin{center}
\shadowrgb{0.8, .8, 1}
\shadowoffset{2.5pt}
\shadowtext{\color{jbul}
\fontsize{13}{13}\selectfont
$\boldsymbol{\bullet}$}
\shadowtext{\color{jour}
\fontsize{15.5}{15.5}\selectfont
\bf Physical Review A 90, 033844-11 (2014)}
\shadowtext{\color{jbul}
\fontsize{13}{13}\selectfont
$\boldsymbol{\bullet}$}
\end{center}

\vspace*{1cm}

\begin{center}
\begin{tabular}{cc}
\begin{minipage}[t]{0,5\textwidth}
{\bf Abstract}.
A detailed analysis of the propagation of laser gaussian beams at critical angles shows in which conditions it is possible to maximize the breaking of symmetry in the  angular distribution  and for which values of the laser wavelength and beam waist is possible to find an analytic formula for angular deviations of the Snell law. For propagation throughout $N$ dielectric blocks and for a full breaking of symmetry, overcoming the well known problem of the infinity at critical angle, a closed expression for the Goos-H\"anchen shift is obtained. The multiple peaks phenomenon clearly represents an additional evidence of the breaking of symmetry in the angular distribution of optical beams. Finally, laser wavelength and beam waist conditions to produce focal effects in the outgoing beam are also briefly discussed.
\end{minipage}
& \begin{minipage}[t]{0,5\textwidth}
{\bf Manoel P. Ara\'ujo}\\
Institute of Physics ``Gleb Wataghin'' \\
State University of Campinas (Brazil)\\
{\color{navy}{{\bf mparaujo@ifi.unicamp.br}}}
\hrule
\vspace*{0.2cm}
{\bf Silv\^ania A. Carvalho}\\
Department of Applied Mathematics\\
State University of Campinas (Brazil)\\
{\color{navy}{{\bf  silalves@ime.unicamp.br}}}
\hrule
\vspace*{0.2cm}
{\bf Stefano De Leo}\\
Department of Applied Mathematics\\
State University of Campinas (Brazil)\\
{\color{navy}{{\bf  deleo@ime.unicamp.br}}}
\end{minipage}
\end{tabular}
\end{center}

\vspace*{2cm}

\begin{center}
\shadowrgb{0.8, 0.8, 1}
\shadowoffset{2pt}
\shadowtext{\color{navy}
{\bf
\begin{tabular}{ll}
I. & INTRODUCTION \\
II. & ASYMMETRICALLY MODELED BEAMS \\
III. & PROPOSING THE BREAKING OF SYMMETRY IN OPTICAL EXPERIMENTS \\
IV.  & SNELL LAW ANGULAR DEVIATION, MULTIPLE PEAKS PHENOMENON\\
 &  AND FOCAL EFFECT \\
V.   & CONCLUSIONS AND OUTLOOKS\\
& \\
& \,[\,18 pages, 2 tables, 6 figures\,]
\end{tabular}
}}
\end{center}

\vspace*{4.5cm}

\begin{flushright}
\shadowrgb{.8, .8, 1}
\shadowoffset{2pt}
\shadowtext{\color{jbul}
\fontsize{11}{11}\selectfont
$\boldsymbol{\bullet}$}
\hspace*{-.2cm}
\shadowtext{
\color{jour}
\fontsize{15.5}{15.5}\selectfont
$\boldsymbol{\Sigma\hspace*{0.06cm}\delta\hspace*{0.035cm}\Lambda}$}
\hspace*{-.35cm}
\shadowtext{\color{jbul}
\fontsize{11}{11}\selectfont
$\boldsymbol{\bullet}$}
\end{flushright}


\newpage


\section*{\normalsize I. INTRODUCTION}

 It is well known that the Fresnel coefficients, which  describe the propagation of optical beams between media with different refractive index,  are useful in studying deviations from geometrical optics\cite{Wolf,Saleh}.
The most important examples are surely represented by the Goos-H\"anchen\cite{GH1947,GH1948,GH1971,GH1972,GH1977,GH1983,GH1986,GH1987,GH2000,GH2013a,GH2013b,GH2014}  and   Imbert-Fedorov\cite{IF1972,IF2004,IF2009,IF2012a,IF2012b,IF2012c}  effects. For total internal reflection, Fresnel coefficients gain an additional phase and this phase is responsible for the transversal shift of linearly and elliptically polarized light with respect to the optical beam path predicted by the Snell law. Nevertheless, these effects do {\em not} modify the angular predictions of geometrical optics. For example, for a dielectric block with parallel sides the outgoing beam is expected to be   parallel to the incoming one. Angular deviations\cite{Ang2002,Ang2006,Ang2007,Ang2009,Ang2013}  from the optical path predicted by the Snell law  are a direct consequence of the breaking of symmetry\cite{AGH2014} in the angular distribution. In this paper, we show how to maximize this breaking of symmetry  and give an {\em analytic formula} for the Snell law angular deviations. Two interesting additional phenomena, i.e. multiple peaks  and focal effect, appear  in the analysis of the outgoing beam. In view of possible experimental investigations,  our study, done for simplicity of presentation for $n=\sqrt{2}$, is then extended to BK7/Fused Silica dielectric blocks and  He-Ne lasers with $\lambda=633\,{\rm nm}$ and beam waists $\mbox{w}_{\0}=100\,\mu{\rm m}$ and $1\,{\rm mm}$.

\section*{\normalsize II. ASYMMETRICALLY MODELED BEAMS}

As anticipated in the introduction, the breaking of symmetry\cite{AGH2014} in the angular distribution of optical beams  plays  a fundamental role in  the angular deviation from the optical  path predicted by the Snell law. In this section, to understand why the breaking of symmetry is responsible for a such  fascinating phenomenon, we briefly discuss a maximal breaking of symmetry for an asymmetrically modeled beam.  The effect of this maximal breaking of symmetry on the peak  and   the position mean value  of  the optical beam  sheds new light on the possibility to realize
an optical experiment.

First of all, let us consider the  {\it symmetric} gaussian angular distribution
 \begin{equation}
 \label{sdis}
g(\theta) = \exp[\,-\,(k\,\mbox{w}_{\0}\,\theta)^{^{2}}\,/\,\,4\,]\,\,,
 \end{equation}
 where $\mbox{w}_{\0}$ is the beam waist of the gaussian laser and $k=2\pi/\lambda$ is the wave number associated to the wavelength $\lambda$. The optical beam, propagating in the $y$-$z$ plane,  is represented by\cite{IF2012b,IF2012c}
 \begin{equation}
 \label{las}
 E(y,z) = E_{\0}\,\frac{k\,\mbox{w}_{\0}}{2\,\sqrt{\pi}}\,\int \mbox{d}\theta\,\,g(\theta)\,\exp[\,i\,k\,(\sin\theta\,y+ \cos \theta\,z)\,]\,\,.
 \end{equation}
For $k\mbox{w}_{\0}\gg 1$, we can develop the sine  and cosine function up to the second order in $\theta$. The  electric field,
\begin{equation}
E(y,z) = \frac{E_{\0}}{\gamma(z)}\,\exp\left\{\,i\,k\,z\,-\,\left[\,\frac{y}{\mbox{w}_{\0}\,{\gamma(z)}}\,\right]^{^{2}}\right\}=
E_{\0}\,e^{i\,k\,z}\,\mathcal{G}(y,z)\,\,,
\end{equation}
 where $\gamma(z)=\sqrt{1+2\,iz/k\,\mbox{w}^{^2}_{\0}}$, thus propagates along the $z$-direction and  manifests a cylindrical symmetry about the direction of propagation. The complex gaussian function $\mathcal{G}(y,z)$  is  solution of the paraxial Helmholtz equation\cite{Wolf,Saleh}
 \begin{equation}
\left(\,\partial_{yy} + 2\,i\,k\,\partial_z\,\right)\, \mathcal{G}(y,z)=0\,\,.
\end{equation}
The optical intensity,
\begin{equation}
\label{gauss}
I(y,z)=|E(y,z)|^{^{2}}= \frac{I_{\0}}{|\gamma(z)|^{^{2}}}\,\,
\exp\left[\,-\,\frac{2\,y^{\2}}{{\rm w}_{\0}^{\2}\,|\gamma(z)|^{^{4}}}\right] =  I_{\0}\,\frac{{\rm w}_{\0}}{{\rm w}(z)}\,\,
\exp\left[\,-\,\frac{2\,y^{\2}}{{\rm w}^{^2}(z)}\right] \,\,,
\end{equation}
is a function  of the axial, $z$, and transversal, $y$, coordinates. The gaussian function  $|\mathcal{G}(y,z)|$ has its peak on the $z$ axis at $y=0$ and its beam width increases with the axial distance $z$ as illustrated in Fig.\,1-a. Due to the fact that the gaussian distribution $g(\theta)$ is a symmetric distribution centered at $\theta=0$,
\begin{equation}
\langle\,  y \,\rangle_{_{|\mathcal{G}|}}=\frac{\displaystyle{\int \mbox{d}y\,y\,|E(y,z)|^{^{2}}}}{ {
\displaystyle{\int  \mbox{d}y\,|E(y,z)|^{^{2}}} }}=\frac{\displaystyle{\int \mbox{d}y\,y\,|\mathcal{G}(y,z)|^{^{2}}}}{ {
\displaystyle{\int  \mbox{d}y\,|\mathcal{G}(y,z)|^{^{2}}} }}=0\,\,.
\end{equation}
 The previous analytical result shows that for symmetric distributions the peak position and transversal mean value coincide and do {\em not} depend on the axial parameter $z$.  The symmetry in the angular distribution $g(\theta)$ is thus responsible for  the well known stationary behavior of  the gaussian laser peak.

To see how   the breaking of symmetry  drastically  changes the previous situation, we model a {\em maximal}
breaking of symmetry by considering the following {\em asymmetric}  angular distribution,
\begin{equation}
\label{adis}
f(\theta) = \left\{\begin{array}{ccrcccl}
0 & \hspace*{.5cm}&  &   & \theta & < & 0\,\,,   \\
\exp[\,-\,(k\,\mbox{w}_{\0}\,\theta)^{^{2}}\,/\,\,4\,] & &  & & \theta & \geq & 0 \,\,.
\end{array}\right.
 \end{equation}
This distribution determines the behavior of the new electric field,
\begin{equation}
\label{nef}
\mathcal{E}(y,z) = \mathcal{E}_{\0}\, \left\{\,1 + \,\mbox{Erf}\, \left[\,i\,y \,/\,\mbox{w}_{\0}\gamma(z)\,\right]\,\right\}\,\mathcal{G}(y,z)=
\mathcal{E}_{\0}\,e^{i\,k\,z}\,\mathcal{F}(y,z)\,\,.
\end{equation}
The asymmetry in the  angular distribution of Eq.(\ref{adis}) is responsible for the axial dependence of the peak position, see Fig.\,1-b. This $z$ dependence is caused by the interference between the gaussian and the error function  which now appears in Eq.(\ref{nef}). The numerical analysis, done for different value of $k{\rm w}_{\0}$ and illustrated in Fig.\,1-c and 1-d, shows a different behavior between the peak position and transversal mean value and confirms the analytical expression
\begin{eqnarray}
\label{ymeanf}
\langle\,  y \,\rangle_{_{|\mathcal{F}|}}&=&\frac{\displaystyle{\int}  \mbox{d}y\,y\,|\mathcal{F}(y,z)|^{^{2}}}{ {\displaystyle{\int}  \mbox{d}y\,|\mathcal{F}(y,z)|^{^{2}}} }
=  \frac{\displaystyle{ -\,\frac{i}{2\,k}\, \int \mbox{d}\theta\,f(\theta)\,e^{-ik\theta^{{\2}}z/2}\frac{\partial}{\partial\theta}\left[f(\theta)\,e^{-ik\theta^{{\2}}z/2}  \right]^{^{*}}
}}{ {\displaystyle{\int}  \mbox{d}\theta\,f^{^{2}}(\theta) }}\, \,\,+\,\,\,{\rm h.c.}
\nonumber \\
& = &  \frac{\displaystyle{\int}  \mbox{d}\theta\,\,\theta\,\,f^{^{2}}(\theta)}{ {\displaystyle{\int}  \mbox{d}\theta\,f^{^{2}}(\theta)} }\,\,z =\frac{\sqrt{2\,/\pi}}{\,\,k\,\mbox{w}_{\0}}\,\,z   \,\,.
\end{eqnarray}
Finally,  the breaking of symmetry in the modeled angular distribution, Eq.(\ref{adis}), generates deviations from the optical path, $y=0$, expected by geometrical optics. The modeled beam now shows the angular deviation
 \begin{equation}
 \label{dev}
 \alpha_{_{\rm max}}=\arctan\left[\frac{\sqrt{2\,/\pi}}{\,\,k\,\mbox{w}_{\0}}   \right]\,\,,
\end{equation}
where the subscript index  has been introduced to recall that this angular deviation is due  to the {\em maximal} breaking of symmetry  introduced to model the gaussian optical beam. This  deviation can be physically understood  by observing that for a symmetric distribution, see $g(\theta)$ in Eq.(\ref{sdis}),   negative and positive angles play the same role  and consequently  their final contribution does not change the propagation of the optical path whose maximum is always centered at $y=0$. In the case of the asymmetric distribution $f(\theta)$ given in Eq.(\ref{adis}) only positive angles contribute to the motion and this generates a {\em maximal} angular deviation which clearly depends on the parameter $k\,\mbox{w}_{\0}$. In the plane wave limit, this deviation tends to zero.

The results presented in this section stimulate to investigate in which situation
 gaussian lasers,  propagating through dielectric blocks, could experience a breaking of symmetry in their angular distributions similar to the modeled breaking of symmetry analyzed in this section. If this happens, the  angular deviation from the optical path predicted by the  Snell law should be  equal to the angle  $\alpha$ given in Eq.\,(\ref{dev}).

\section*{\normalsize III. PROPOSING THE BREAKING OF SYMMETRY IN OPTICAL EXPERIMENTS}

In this section, we treat the general problem of the transmission of a gaussian optical beam through a dielectric block and study how to realize the breaking of symmetry which allows to reproduce the effects discussed in the previous section. What done in this section is only a {\em proposal} to observe the breaking of symmetry in real optical experiments and  see in which circumstances it is possible to reproduce the maximal angular deviation
 $\alpha_{_{\rm max}}$ of Eq.(\ref{dev}). In this proposal, we do not take into account cumulative dissipation effects. Imperfections such as misalignment of the dielectric surfaces will be discussed in the final section.

The optical beam represented by the electric field of Eq.(\ref{las}) moves from its source $S$ to the left interface of the dielectric block  along the $z$-axis, see  Fig.\,2-a. The $\widetilde{z}$ and $z_*$ directions  represent respectively the left/right and up/down stratifications of the dielectric block. By observing that
\begin{equation}
\left( \begin{array}{c}
y \\
z\end{array} \right)  = \left( \begin{array}{rr}
\cos\theta_{\0} & -\sin\theta_{\0} \\
\sin\theta_{\0} & \cos\theta_{\0} \end{array} \right)\, \left( \begin{array}{c}
\widetilde{y} \\
\widetilde{z} \end{array} \right) \,\,,
\label{eq:rot}
\end{equation}
we can immediately rewrite the incoming electric field  in terms of the new axes $\widetilde{y}$ and $\widetilde{z}$,
 \begin{eqnarray}
 E_{_{\rm inc}}(y,z) &=& E_{\0}\,\frac{k\,\mbox{w}_{\0}}{2\,\sqrt{\pi}}\,\int \mbox{d}\theta\,\,g(\theta)\,\exp[\,i\,k\,(\,\sin\theta\,y+ \cos \theta\,z\,)\,]\nonumber\\
 & = &
 E_{\0}\,\frac{k\,\mbox{w}_{\0}}{2\,\sqrt{\pi}}\,\int \mbox{d}\theta\,\,g(\theta)\,\exp\{\,i\,k\,[\,\sin(\theta+\theta_{\0})\,\widetilde{y}+ \cos (\theta+\theta_{\0})\,\widetilde{z}\,]\,\}\nonumber\\
  & = & E_{\0}\,\frac{k\,\mbox{w}_{\0}}{2\,\sqrt{\pi}}\,\int \mbox{d}\theta\,\,g(\theta-\theta_{\0})\,
  \exp[\,i\,k\,(\,\sin\theta\,\widetilde{y}+ \cos \theta\,\widetilde{z}\,)\,]\,\,.
 \end{eqnarray}
At the first (left) and last (right) interface, $\sin \theta=n\,\sin \psi$, see the dielectric block of Fig.\,2. In terms of these angles, the transmission Fresnel coefficients for $s$ polarized waves are given by\cite{Wolf,Saleh}
\begin{equation}
\label{tlr}
\left\{\,T^{^{[s]}}_{_{\rm left}}\,,\, T^{^{[s]}}_{_{\rm right}} \,  \right\} = \left\{\,
\frac{2\,\cos \theta}{\cos\theta + n\cos \psi}\,\,e^{i\,\phi_{{\rm left}}}\,,\,
\frac{2\,n\cos \psi\,e^{i\,\phi_{{\rm right}}}}{\cos\theta + n\cos \psi}\,\,\,e^{i\,\phi_{{\rm right}}}
\,\right\}\,\,,
\end{equation}
where
\[\phi_{_{\rm left}} = k\,(\cos\theta-n\cos\psi)\, \overline{S\widetilde{D}}\,\,\,\,\,\,\,\mbox{and}
\,\,\,\,\,\,\, \phi_{_{\rm right}} = k\,(n\cos\psi- \cos\theta)\, \left(\,\overline{S\widetilde{D}} + \frac{\overline{BC}}{\sqrt{2}}\,\right) \,\,.    \]
The phase which appears in  the Fresnel coefficients contains information on the point in which the beam
encounters the air/dielectric (dielectric/air) interface and it is obviously equal for $s$ and $p$ polarized waves\cite{Phase2011,Phase2013a,Phase2013b}.  At the second (down) and third (up) interface, observing that $\varphi =\psi + \frac{\pi}{4}$ (see Fig.\,2-a), the reflection Fresnel coefficients read
\begin{equation}
\label{rud}
\left\{\,R^{^{[s]}}_{_{\rm down}}\,,\,R^{^{[s]}}_{_{\rm up}} \,  \right\} =  \frac{n\cos \varphi - \sqrt{1-n^{^{2}}\sin^{\2}\varphi}} {n\cos \varphi + \sqrt{1-n^{^{2}}\sin^{\2}\varphi}} \,\left\{\,
\,e^{i\,\phi_{{\rm down}}}\,,\, e^{i\,\phi_{{\rm up}}}    \,\right\}\,\,,
\end{equation}
where
\[\phi_{_{\rm down}} = 2\,k\,n\cos\varphi\, \overline{SD_*}\,\,\,\,\,\,\,\mbox{and}
\,\,\,\,\,\,\, \phi_{_{\rm up}} = 2\,k\,n\cos\varphi \, \left(\, \frac{\overline{AB}}{\sqrt{2}} -  \overline{SD_*} \, \right) \,\,.    \]
The total transmission coefficient for $s$ polarized waves which propagate through the dielectric block sketched in
Fig.\,2-a  is then obtained by multiplying the Fresnel coefficients given in Eqs.\,(\ref{tlr}-\ref{rud}),
\begin{equation}
T^{^{[s]}}(\theta) = \frac{4\,n\cos\psi \cos \theta}{\left(\,\cos\theta + n\cos \psi\,\right)^{^{2}}}
\left(\,\frac{n\cos \varphi - \sqrt{1-n^{^{2}}\sin^{\2}\varphi}} {n\cos \varphi + \sqrt{1-n^{^{2}}\sin^{\2}\varphi}}
\,\right)^{^{2}}\,e^{i\,\Phi_{\rm Snell}}\,\,,
\label{TS}
\end{equation}
where
\[ \Phi_{_{\rm Snell}}=  k\,\left[\,
\sqrt{2}\,n\cos\varphi \, \overline{AB} +  (n\cos\psi- \cos\theta)\, \frac{\overline{BC}}{\sqrt{2}}\,\right]\,\,.\]
In a similar way, we can immediately obtain the transmission coefficient for $p$ polarized waves\cite{Phase2013a,Phase2013b},
\begin{equation}
T^{^{[p]}}(\theta) = \frac{4\,n\cos\psi \cos \theta}{\left(\,n\cos\theta + \cos \psi\,\right)^{^{2}}}
\left(\,\frac{\cos \varphi - n\sqrt{1-n^{^{2}}\sin^{\2}\varphi}} {\cos \varphi + n\sqrt{1-n^{^{2}}\sin^{\2}\varphi}}
\,\right)^{^{2}}\,e^{i\,\Phi_{\rm Snell}}\,\,.
\label{TP}
\end{equation}
Before to discuss the effect of the transmission coefficient on the angular gaussian distribution, $g(\theta-\theta_{\0})$, let us spend some time for analyzing the phase, $\Phi_{\rm Snell}$,  which appears in the transmission coefficient. The stationary phase approximation\cite{SPM1955,SPM1973,SPM1975}, which is a basic principle of asymptotic analysis based on the cancellation of sinusoids with rapidly varying phase, allows to obtain a prediction  of the beam peak position by imposing
\[ \left[\,\frac{\partial}{\partial\theta}\,\left(\,k\,\sin \theta\,\widetilde{y}_{_{\rm out}} + k\,\cos \theta\,\widetilde{z}_{_{\rm out}} +\,
\Phi_{_{\rm Snell}}\,\right)\,\right]_{\theta=\theta_{\0}} = 0\,\,.  \]
This stationary constraint implies
\begin{eqnarray}
\cos\theta_{\0}\,\widetilde{y}_{_{out}} - \sin\theta_{\0}\,\widetilde{z}_{_{\rm out}}  &=&
\sqrt{2}\,\,\sin\varphi_{\0} \,\frac{\cos\theta_{\0}}{\cos\psi_{\0}}\,\, \overline{AB}  +  (\,\sin\psi_{\0}\,\frac{\cos\theta_{\0}}{\cos\psi_{\0}} - \sin\theta_{\0}\,)\,\,   \frac{\overline{BC}}{\sqrt{2}}  \nonumber \\
 & = & \underbrace{\cos\theta_{\0}\, \left[\, (\,1+\tan\psi_{\0}\,)\,\overline{AB}  + (\,\tan\psi_{\0} - \tan\theta_{\0}\,)\, \frac{\overline{BC}}{\sqrt{2}} \,\right]}_{\mbox{$d_{_{\rm Snell}}$}} \,\,.
\end{eqnarray}
This reproduces the well known transversal shift obtained in geometrical optics by using the Snell law. With respect to the incoming optical beam, which is centered at $y=0$, the center of the outgoing beam is then shifted at $y=d_{_{\rm Snell}}$. To ensure that for the dielectric structure illustrated  in Fig.\,2-c, we have $2\,N$ internal reflections, we must impose that, in each block,  incoming and outgoing beams have the same $z_*$ component, this implies
\begin{equation}
\label{cond}
\overline{BC} =\sqrt{2}\,\tan \varphi_{\0}\,\overline{AB} \,\,.
\end{equation}
In this case, the propagation of the optical beam through  $N$ dielectric blocks  is characterized by $2\,N$ internal reflections. For an elongated prism of side $N\,\overline{BC}$,  the transmission coefficients for $s$ and $p$ polarized waves  are then  given by
\begin{equation}
T^{^{[s]}}_{_{N}}(\theta) = \frac{4\,n\cos\psi \cos \theta}{\left(\,\cos\theta + n\cos \psi\,\right)^{^{2}}}
\left(\,\frac{n\cos \varphi - \sqrt{1-n^{^{2}}\sin^{\2}\varphi}} {n\cos \varphi + \sqrt{1-n^{^{2}}\sin^{\2}\varphi}}
\,\right)^{^{2\,N}}\,e^{i\,N\,\Phi_{\rm Snell}}\,\,,
\end{equation}
and
\begin{equation}
T^{^{[p]}}_{_{N}}(\theta) = \frac{4\,n\cos\psi \cos \theta}{\left(\,n\cos\theta + \cos \psi\,\right)^{^{2}}}
\left(\,\frac{\cos \varphi - n\sqrt{1-n^{^{2}}\sin^{\2}\varphi}} {\cos \varphi + n\sqrt{1-n^{^{2}}\sin^{\2}\varphi}}
\,\right)^{^{2\,N}}\,e^{i\,N\,\Phi_{\rm Snell}}\,\,.
\end{equation}
For incidence angles lesser than the critical angle,
\[  \theta < \theta_c = \arcsin\left\{\,n\,\sin\left[ \arcsin\left(\frac{1}{n}\right) - \frac{\pi}{4}     \right]\,\right\}\,\,,\]
the outgoing  optical beam,
\begin{eqnarray}
 E_{_T}^{^{[s,p]}}(y,z)
  &= & E_{\0}\,\frac{k\,\mbox{w}_{\0}}{2\,\sqrt{\pi}}\,\int  \mbox{d}\theta\,\,T^{^{[s,p]}}_{_{N}}(\theta)\, g(\theta-\theta_{\0})\,
  \exp[\,i\,k\,(\,\sin\theta\,\widetilde{y}+ \cos \theta\,\widetilde{z}\,)\,]\nonumber \\
   & = & E_{\0}\,\frac{k\,\mbox{w}_{\0}}{2\,\sqrt{\pi}}\,\int \mbox{d}\theta\,\,g_{_{T}}^{^{[s,p]}}(\theta;\theta_{\0})\,
  \exp\{\,i\,k\,[\,\sin(\theta-\theta_{\0})\,y+ \cos (\theta-\theta_{\0})\,z\,]\,\}\,\,,
 \end{eqnarray}
propagates  parallel to the $z$-axis and with its peak located at
\begin{equation}
y_{_{\rm Snell}} = N\,d_{_{\rm Snell}} = N\,(\,\cos\theta_{\0} -\sin\theta_{\0}\,)\,\tan\varphi_{\0}\,\,\overline{AB}\,\,,
\end{equation}
as expected from the ray optics.
For incidence angles greater than the critical angle, we find $\sin\varphi>1$ and the optical beam gains an {\em additional} phase,
\begin{equation}
N\,\Phi_{_{\rm GH}}^{^{[s,p]}}\,\,,
\end{equation}
where
\begin{equation}
\label{ghphase}
\Phi_{_{\rm GH}}^{^{[s,p]}} =
\left\{
\begin{array}{ccl}
-\,4\,\arctan \sqrt{(n^{^{2}}\sin^{\2}\varphi\,-\,1)\, \,/\,(n\cos \varphi)^{^{2}}} & &
\,\,\,\,[\,s\,\, {\rm polarization}\,]\,\,,\\
-\,4\,\arctan \sqrt{n^{^{2}}(n^{^{2}}\sin^{\2}\varphi\,-\,1)\, \,/\,\cos^{^{2}} \varphi} & &
\,\,\,\,[\,p\,\, {\rm polarization}\,]\,\,.
\end{array}
\right.
\end{equation}
For linearly polarized light, this new phase is responsible  for the  Goos-H\"anchen shift. Shift  experimentally observed in 1947\cite{GH1947} and for which,  one year later,  Artman\cite{GH1948} proposed an analytical expression. The Artman formulas, valid for incidence angle greater than the critical angle, have been recently generalized for incidence at critical angle\cite{GH2014}. Notwithstanding the interesting nuances involved in the study of the Goos-Hanch\"en shift, what we aim to discuss in detail in this paper is the {\em angular deviation} from the  optical path predicted by the Snell law.

The angular deviation $\alpha_{_{\rm max}}$,  given in Eq.\,(\ref{dev}),  is  due to the {\em maximal}\, breaking of symmetry  modeled in section II, see Eq.\,(\ref{adis}). In the dielectric structure illustrated in Fig.\,2-c
(observe that in a real optical experiment this structure can be reproduced by a single elongated prism of side $N\,\overline{BC}$), the optical beam experiences $2\,N$  internal reflections and this will play a fundamental role in reproducing,  for incidence at critical angle,   the maximal breaking of symmetry presented  in section II by a modeled example. Indeed, for incidence at critical angle, the angular distribution $g_{_{T}}^{^{[s,p]}}(\theta;\theta_{c})$ centered at $\theta=\theta_{c}$  suffers,  at each up and down interface, a partial transmission for $\theta<\theta_c$ and a total reflection for  $\theta>\theta_c$.  Consequently, by increasing  the number of internal reflections we contribute to increase the breaking of symmetry. For few blocks, the real optical experiment is very different from the modeled case presented in section II. Nevertheless, for $N\gg 1$, we improve the breaking of symmetry and can simulate the maximal breaking of symmetry discussed in section II. From  Fig.\,3, where we plot the modulus of the transmitted angular distribution  $g_{_{T}}^{^{[s,p]}}(\theta;\theta_{c})$, we can immediately see that the breaking of symmetry is optimized not only by increasing the number of blocks but also using $p$ polarized waves and/or decreasing the value of the beam waist. As shown in Fig.\,3-c, for $k{\rm w}_{\0}=10^{^3}$  (which for He-Ne laser with $\lambda=633\,{\rm nm}$ means ${\rm w}_{\0}\approx 100\,\mu{\rm m}$),  $N=50$ and $p$ polarized waves, we {\em perfectly} reproduce  the modeled breaking of symmetry  illustrated in section II. By increasing the number of blocks or equivalently the side of the elongated prism, we can always reach the maximal breaking of symmetry (\ref{adis}). It is important to be observed here that such a distribution leads to the {\em maximal} angular deviation. For incidence not at critical angle or in the presence of misalignment at the dielectric surfaces the angular deviation decreases (see discussion at the end of the final section).

\section*{\normalsize IV. SNELL LAW ANGULAR DEVIATION, MULTIPLE PEAKS PHENOMENON AND FOCAL EFFECT}
As observed in the previous section, it is possible to reproduce in a real optical experiment the modeled breaking of symmetry introduced in section II. The preferred incidence angle is $\theta_{\0}=\theta_c$. In this case, for an appropriate choice of the number of dielectric blocks ($N=50$)  and of the laser beam waist ($k{\rm w}_{\0}=10^{^3}$), it is possible to take the following approximation
\begin{equation}
g_{_{T}}^{^{[s,p]}}(\theta;\theta_{c}) = \left|g_{_{T}}^{^{[s,p]}}(\theta;\theta_{c})\right|\,e^{i\,N\,(\,\Phi_{_{\rm Snell}}\,\,+\,\,\Phi_{_{\rm GH}}^{^{[s,p]}}\,)}
\approx f(\theta-\theta_c)\, e^{i\,N\,(\,\Phi_{_{\rm Snell}}\,\,+\,\,\Phi_{_{\rm GH}}^{^{[s,p]}}\,)}\,\,.
\end{equation}
The transversal mean value for the outgoing beam is then given  by
\begin{eqnarray}
\label{ymeanTra1}
\langle\,  y \,\rangle_{_{T,c}}^{^{[s,p]}}&=&
  \frac{\displaystyle{ -\,\frac{i}{2\,k}\, \int \mbox{d}\theta\,g_{_{T}}^{^{[s,p]}}(\theta;\theta_{c})\,e^{-ik(\theta-\theta_c)^{{\2}}z/2}
\frac{\partial}{\partial\theta}\left[ g_{_{T}}^{^{[s,p]}}(\theta;\theta_c)  \,e^{-ik(\theta-\theta_c)^{{\2}}z/2}  \right]^{^{*}}
}}{ {\displaystyle{\int}  \mbox{d}\theta\,\left|g_{_{T}}^{^{[s,p]}}(\theta;\theta_{c})\right|^{^{2}} }}\, \,\,+\,\,\,{\rm h.c.}
\nonumber \\
 & = & \frac{\displaystyle{ \int \mbox{d}\theta\,  \left[\,
-\,\frac{N}{k}\,\frac{\partial}{\partial\theta}\,\left( \,\Phi_{_{{\rm Snell}}}\,\,+\,\,\Phi_{_{{\rm GH}}}^{^{[s,p]}}\right)  \,+\, (\theta-\theta_{c})\,\,z\right]
\,\,f^{^{2}}(\theta-\theta_c)}}{\displaystyle{ \int \mbox{d}\theta
\,\,f^{^{2}}(\theta-\theta_c)}}\nonumber \\
 & = & y_{_{{\rm Snell},c}}\, +\,\,  y_{_{{\rm GH},c}}^{^{[s,p]}}\, +\, \frac{\sqrt{2\,/\pi}}{\,\,k\,\mbox{w}_{\0}}\,\,z   \,\,.
\end{eqnarray}
The Snell law angular deviation $\alpha$, see Eq.\,(\ref{dev}) and Fig.\,2-b, obtained in section II for a modeled  breaking of symmetry can be now reproduced in a real optical experiment.  For a partial breaking of symmetry the angular deviation is obviously reduced and a numerical calculation is needed to estimate  such a deviation, see Fig.\,4. The peak position and the transversal component mean value, plotted in Fig.\,4 for $n=\sqrt{2}$ which has been chosen because a dielectric block with such a refractive index has a critical angle $\theta_c=0$ ($\varphi_c=\pi/4$), have been also calculated for dielectric Fused Silica ($n=1.457$) and BK7 ($n=1.515$) blocks,
\[ \left\{\,n\,,\,\frac{180^{^{o}}\,\theta_c}{\pi}\,,\,  \frac{180^{^{o}}\,\varphi_c}{\pi}  \,\right\} \,\,=\,\, \left\{\, \sqrt{2} \,,\,0^{^{o}} \,,\,45^{^{o}}\,\right\}\,\,,\,\,\,\, \left\{\, 1.457 \,,\,-\,2.42^{^{o}} \,,\, 43.34^{^{o}}\,\right\}\,\,,\,\,\,\,  \left\{\,1.515\,,\, -\,5.60^{^{o}}\,,\,41.31^{^{o}}\,\right\}\,\,, \]
see Tab.\,1.  It is important to be observed that increasing the number of blocks we reach a maximal  breaking of symmetry. In the Snell and Goos-H\"anchen shifts, we find a {\em linear} dependence on the blocks number,
\begin{eqnarray}
y_{_{{\rm Snell},c}} & = &   N\,(\,\cos\theta_{c} -\sin\theta_{c}\,)\,\tan\varphi_{c}\,\,\overline{AB}\,\,,
\nonumber \\
 & = &   N\,\frac{\sqrt{2-n^{\2}+2\,\sqrt{n^{\2}-1}} - 1 + \sqrt{n^{\2}-1}}{\sqrt{2\,(n^{\2}-1)}}\,\,\overline{AB}\,\,,\nonumber\\
 & = & N\,\delta_{_{{\rm Snell},c}}\,\overline{AB}\,\,,
\end{eqnarray}
and
\begin{eqnarray}
\label{GHcri}
\left\{\,y_{_{{\rm GH},c}}^{^{[s]}}\,,\,  y_{_{{\rm GH},c}}^{^{[p]}}  \right\}& = &
N\,\left\{\,1\,,\,n^{\2}\,\right\}\,\frac{ \displaystyle{\frac{4}{k}\, \int \mbox{d}\theta\,
\sqrt{\frac{\cos\theta_c}{n\,\cos\psi_c\,(\,n^{^{2}}-2\,\sin^{^2}\theta_c\,)\,\,(\theta-\theta_c)}}
\,\,\,f^{^{2}}(\theta-\theta_c)}}{\displaystyle{ \int \mbox{d}\theta
\,\,f^{^{2}}(\theta-\theta_c)}}\nonumber \\
 & = & N\,\left\{\,1\,,\,n^{\2}\,\right\}\,\frac{4\,\Gamma(1/4)}{\sqrt{\pi\sqrt{2}}}\,\,\left[\frac{2\,-\,n^{^{2}}+\,2\,
\sqrt{n^{^{2}}-1}}{4\,(n^{^{2}}-1)\,\left(   n^{^{2}}+\,2\,\sqrt{n^{^{2}}-1}\,\right)} \right]^{^{1/4}}  \,\sqrt{\frac{{\rm w}_{\0}}{k}}\nonumber\\
 & = &  N\,\left\{\,1\,,\,n^{\2}\,\right\}\, \delta_{_{{\rm GH},c}} \, \sqrt{\frac{{\rm w}_{\0}}{k}}  \,\,.
\end{eqnarray}
Note that the divergence at critical angle is removed by the previous integration. Consequently, for  a maximal breaking of symmetry, we find an analytical expression for the Goos-H\"anchen shift at critical angle. Observing that
\[ \left\{\,n\,,\,\delta_{_{{\rm Snell},c}}\,,\, \delta_{_{{\rm GH},c}}\,\right\} \,\,=\,\,
\left\{\, \sqrt{2} \,,\,1 \,,\,4.091\,\right\}\,\,\,,\,\,\,\,\,
\left\{\, 1.457 \,,\,0.983 \,,\,3.915\,\right\}\,\,\,,\,\,\,\,\,
\left\{\, 1.515 \,,\,0.960 \,,\,3.700\,\right\}\,\,,
\]
in the $N$ blocks dielectric structure of Fig.\,2-c, we have a Snell shift proportional to $N\,\overline{AB}$
and  an amplification of the standard Goos-H\"anchen shift ($\sim \lambda$) given by $N\,\sqrt{k\mbox{w}_{\0}}$. The numerical analysis done in Fig.\,4 and Tab.\,1
confirms this amplification.

The multiple peaks phenomenon observed in Fig.\,5 is a clear evidence of the breaking of symmetry in the angular distribution. In the optical beam, the negative angular contributions  are suppressed  if we increase the number of blocks.  This implies only positive angular contributions in the spreading of the optical beam and consequently the multiple peaks phenomenon. As can be seen in Fig.\,5, this phenomenon is amplified  not only by increasing the number of blocks but also by using $p$-polarized waves. Observe that the phenomenon is more evident when the spreading of the optical beam is clearly visible. In Fig.\,6, it is present another interesting phenomenon. The focal effect in the outgoing beam is a consequence of the second order contribution of the optical phase which is responsible for the spreading of the beam. The numerical analysis, see Fig.\,6-f and Tab.\,2, shows an increasing value of the maximum of the outgoing electrical field and this is a clear evidence of a focalization of the beam. From the data presented in Tab.\,2, we can estimate the axial point of maximal focalization.

\section*{\normalsize V. CONCLUSIONS AND OUTLOOKS}

The connection between quantum mechanics and optics\cite{GH2013a} and the possibility to realize optical experiments to reproduce quantum effects\cite{GH2013b} makes optics an interesting subject of study to investigate  the most diversified phenomena, from the Goos-H\"anchen and Imbert-Federov shifts to the frustrated total internal reflection\cite{FTIR1,FTIR2,FTIR3,FTIR4} and resonant photon tunnelling\cite{RLT}.
In this paper, starting from a modeled symmetry breaking (section II) we have shown {\em how} to reproduce the {\em maximal} breaking of symmetry in the angular distribution of laser beams by an optical structure composed by
$N$ dielectric blocks. This structure can be realized in real optical experiment by a single elongated prism.
The breaking of symmetry causes an {\em angular modification} for the optical path predicted by the Snell law. The outgoing beam is no longer parallel to the incoming one as expected from the Snell law.  Our analysis   shows that the maximal angular deviation is obtained for a gaussian He-Ne laser with $\lambda=633$\,nm and beam waist w$_{\0}=100\,\mu$m  by using $p$-polarized waves and a dielectric structure with $50$ blocks (see Fig.\,2-c). In this case, we should find an angular deviation
\begin{equation}
\alpha_{_{\rm max}} =\arctan \left[ \, \frac{\displaystyle{ \int \mbox{d}\theta\, \, (\theta-\theta_{_{c}})\,\,f^{^{2}}(\theta-\theta_{_{c}})}}{\displaystyle{ \int \mbox{d}\theta
\,\,f^{^{2}}(\theta-\theta_{_{c}})}}\,\right] =  \arctan \left[  \frac{\sqrt{2/\pi}}{k\,{\rm w}_{\0}} \,  \right]  \approx \frac{\sqrt{2/\pi}}{k\,{\rm w}_{\0}} \approx 0.05^{^{o}}\,\,\frac{\pi}{180^{^{o}}} \,\,,
\end{equation}
which does  not depend on the refractive index $n$ of the dielectric blocks and does  not change if we increase the blocks number because we have reached the maximal breaking of symmetry for $N=50$.  This prediction can be tested in real optical experiments by using different dielectric blocks, for example Fused Silica and BK7 (see Table 1). Nevertheless, the previous formula does {\em not} take into account cumulative dissipations and imperfections in the prism such as misalignment of its surfaces. A phenomenological way to include the misalignment effect in the angular deviation is to consider the following distribution
\begin{equation}
f_{_{\rm mis}}(\theta-\theta_{_c}) = \left\{\begin{array}{ccrcccl}
0 & \hspace*{.5cm}&  &   & \theta-\theta_{_{c}} & < & \theta_{_{\rm mis}}\,\,,   \\
\exp[\,-\,(k\,\mbox{w}_{\0}\,\theta)^{^{2}}\,/\,\,4\,] & &  & & \theta-\theta_{_{c}} & \geq & \theta_{_{\rm mis}} \,\,,
\end{array}\right.
 \end{equation}
where the angle $\theta_{_{\rm mis}}=\arcsin \left[ n\,\sin \left(\varphi_{_{\rm mis}} -\frac{\pi}{4}\right)\right]$ is introduced to include misalignment effects. Such effects can be simulated by  observing that the surfaces misalignment can be simulated by changing the internal angle from  $\varphi_{_{c}}$ to   $\varphi_{_{c}}+\varphi_{_{\rm mis}}$. In this case, the angular deviation is given by
\begin{equation}
\label{mis}
\alpha_{_{\rm mis}} =\arctan \left[ \, \frac{\displaystyle{ \int \mbox{d}\theta\, \, \theta\,\,f_{_{\rm mis}}^{^{2}}(\theta-\theta_{_{c}})}}{\displaystyle{ \int \mbox{d}\theta
\,\,f_{_{\rm mis}}^{^{2}}(\theta-\theta_{_{c}})}}\,\right] \approx
\displaystyle{\frac{\exp[\,-\,(k\,\mbox{w}_{\0}\,\theta_{_{\rm mis}})^{^{2}}\,/\,\,2\,]}{\mbox{Erfc}\left[ \,
k\,\mbox{w}_{\0}\,\theta_{_{\rm mis}}\,/\,\sqrt{2} \,\right]}}\,\, \alpha_{_{\rm max}}\,\,.
\end{equation}
The possibility to realize a {\em maximal} breaking of symmetry and to make a prediction on the angular deviation of the Snell law surely represents the main objective of our investigation. The study done in this article, overcoming the infinity at critical angle by the integration of Eq.\,(\ref{GHcri}), allows also to find a closed-form expression for the Goos-H\"anchen shift. The prediction is in excellent agreement with our numerical calculation. Finally, but not less important, two additional phenomena appears in the presence of the symmetry breaking, namely the multiple peaks phenomenon and the focal effect. We hope that the analysis presented in this work stimulates optical experiments to confirm the angular deviation $\alpha$ (Fig.\,4) , the multiple peaks (Fig.\,5) and the focalization (Fig.\,6) in the outgoing beam.

In a forthcoming paper, we aim to extend the investigation of the symmetry breaking  done in this work for gaussian
angular distributions by analyzing the effect of the breaking of symmetry in  hermite and laguerre gaussian optical beams\cite{F1}.

\vspace*{2cm}

\noindent
\textbf{\small ACKNOWLEDGMENTS.} The authors gratefully acknowledge  Capes (M.P.A.),  CNPq (S.D.L.), and  Fapesp (S.A.C.) for the the financial support and the referee for his comments and suggestions, in particular for drawing the attention to the  misalignment effect which  stimulated the discussion which led to Eq.(\ref{mis}).


\newpage

\begin{center}
\begin{table}[ht]
 \caption{{\bf Snell law angular deviation for BK7 and Fused Silica dielectric blocks.} Numerical peak position, $y_{_{{\rm max},T} }$, and transversal mean value, $\langle y \rangle_{_{T}}$, of the transmitted beam at critical angle  are listed, for $s$ and $p$ polarized waves, for different refractive index as function of the number of blocks, $N$, and  for fixed  beam waist/wavelength ratio, $k{\rm w}_{\0}$, and axial parameter, $z$. From the table, we immediately see that by increasing the number of blocks (and consequently  optimizing the breaking of symmetry) we increase the angular deviation of the Snell law.}
\renewcommand{\arraystretch}{1.4}
 \hspace*{-0.5cm} \colorbox{gray!10}{
  \begin{tabular}{ |c||p{1.5cm} |p{1.5cm} |p{1.5cm}||p{1.5cm} | p{1.5cm} |p{1.5cm}||p{1.2cm}|}
\hline
&\multicolumn{3}{c||}{ \multirow{2}{*}{\Huge\bf$\boldsymbol{ \frac{y_{_{\mbox{\normalsize max,$T$} }}-\, y_{_{\mbox{\normalsize Snell}  }}}{d_{_{ \mbox{\normalsize Snell}  }}}}$}} & \multicolumn{3}{c||}{ \multirow{2}{*}{\Huge \bf $\boldsymbol{\frac{\langle\, y \,- \, y_{_{\mbox{\normalsize Snell} }}\rangle_{_{ \mbox{\normalsize $T$}}}}{d_{_{ \mbox{\normalsize Snell}}}}}$}} &\\
& \multicolumn{3}{c||}{}& \multicolumn{3}{c||}{}&\\
& \multicolumn{3}{c||}{}& \multicolumn{3}{c||}{}&\\
\hline
\hline
\diagbox[width=2.6cm,height=1.1cm]{\hspace{0.8cm}\Large$ \boldsymbol{n}$}{\hspace{-1cm}\Large$\boldsymbol{N}$} & \hspace{0.45cm}\Large$\textcolor{orange} {\boldsymbol{10}}$  &\hspace{0.55cm}\Large$\textcolor{cyan} {\boldsymbol{30}}$  &\hspace{0.5cm}\Large$\textcolor{blue} {\boldsymbol{50}}$ &\hspace{0.45cm}\Large$\textcolor{orange} {\boldsymbol{10}}$ &\hspace{0.4cm}\Large $\textcolor{cyan} {\boldsymbol{30}}$ &\hspace{0.5cm}\Large$\textcolor{blue} {\boldsymbol{50}}$ & \\\hline
 \hline
\Large$ \boldsymbol{\sqrt{2}}$ &  \hspace{0.3cm}\textcolor{orange} {0.050 } &\hspace{0.3cm} \textcolor{cyan} {0.080}  & \hspace{0.3cm}\textcolor{blue} {0.108}  & \hspace{0.2cm} \textcolor{orange} {0.053} & \hspace{0.3cm}\textcolor{cyan} {0.079}  &\hspace{0.3cm}\textcolor{blue} { 0.104} &\vspace{-0.8cm}\hspace{0.4cm}\multirow{3}{*}{\rotatebox{-90}{\Large\textbf{ s\,-\,pol}}}  \\

 \cline{1-7}
\Large$\boldsymbol{1.457} $ &  \hspace{0.3cm}\textcolor{orange} {0.049}  & \hspace{0.4cm}\textcolor{cyan} {0.079}  &\hspace{0.3cm}\textcolor{blue} { 0.105} & \hspace{0.3cm}\textcolor{orange} {0.052}  &\hspace{0.2cm} \textcolor{cyan} {0.077}  & \hspace{0.3cm}\textcolor{blue} {0.102}&\\
  \cline{1-7}
\Large$\boldsymbol{1.515}$  &  \hspace{0.3cm}\textcolor{orange} {0.047}  & \hspace{0.4cm}\textcolor{cyan} {0.077}  & \hspace{0.3cm}\textcolor{blue} {0.102}   & \hspace{0.3cm}\textcolor{orange} {0.051}  & \hspace{0.3cm}\textcolor{cyan} {0.075} &\hspace{0.3cm}\textcolor{blue} { 0.099}& \\\hline
  \hline
  \Large $\boldsymbol{\sqrt{2}}$  &  \hspace{0.3cm}\textcolor{orange} {0.063}  &\hspace{0.4cm}\textcolor{cyan} { 0.112}  & \hspace{0.3cm}\textcolor{blue} {0.156} & \hspace{0.3cm}\textcolor{orange} {0.066}  & \hspace{0.3cm}\textcolor{cyan} {0.117}  &\hspace{0.3cm}\textcolor{blue} { 0.170} & \vspace{-0.8cm}\hspace{0.4cm}\multirow{3}{*}{\rotatebox{-90}{\Large\textbf{ p\,-\,pol}}} \\
    \cline{1-7}
\Large$\boldsymbol{1.457}$  &  \hspace{0.3cm}\textcolor{orange} {0.064}  & \hspace{0.4cm}\textcolor{cyan} {0.113}  & \hspace{0.3cm}\textcolor{blue} {0.158}  & \hspace{0.3cm}\textcolor{orange} {0.066}  & \hspace{0.3cm}\textcolor{cyan} {0.119}  & \hspace{0.3cm}\textcolor{blue} {0.171} &\\
  \cline{1-7}
\Large$ \boldsymbol{1.515}$  &  \hspace{0.3cm}\textcolor{orange} {0.064}  & \hspace{0.4cm}\textcolor{cyan} {0.114} & \hspace{0.3cm}\textcolor{blue} {0.159} & \hspace{0.3cm}\textcolor{orange} {0.067} & \hspace{0.3cm}\textcolor{cyan} {0.120}& \hspace{0.3cm}\textcolor{blue} {0.174 } &\\\hline
\hline
&\multicolumn{6}{c||}{ \Large\bf $\bullet\,\,\,\,\boldsymbol{k\mbox{w}_{_0}=10^{^3}}\,\,\,\,\bullet\,\,\,\,
\boldsymbol{\theta_{0}=\theta_{c}}\,\,\,\,\bullet\,\,\,\,\boldsymbol{z=50\,d_{_{\rm Snell}}}\,\,\,\,\bullet$ }&\\\hline
    \end{tabular}
}
\end{table}
\end{center}
\newpage
\begin{center}
\begin{table}[ht]
 \caption{\textbf{Focal effect for BK7 and Fused Silica dielectric blocks.} Numerical maximum  of the outgoing electrical field at critical angle is listed,  for $p$ polarized waves, for different refractive index as function of the axial parameter $z$ and for fixed  beam waist/wavelength ratio, $k{\rm w}_{\0}$, and blocks number, $N$. From the table, we clear see the focalization near
 $z=10^{^{3}}\,d_{_{\rm Snell}}$.}
\renewcommand{\arraystretch}{1.8}
 \setlength{\tabcolsep}{8pt}
\captionsetup[table]{skip=10pt}
\colorbox{gray!10}{
   \begin{tabular}{|c||p{1.5cm}|p{1.5cm}|p{1.5cm}|p{1.5cm}|p{1.4cm}||p{1.2cm}|}
\hline
&\multicolumn{5}{c||}{\multirow{2}{*}{\Huge\bf $\boldsymbol{\big|E_{_{\mbox{\normalsize $T$}}}\left(y,z\right)/E_{_{\mbox{\normalsize $0$}}}\big|_{_{\mbox{\normalsize  max}}}}$}} &\\
&\multicolumn{5}{c||}{} &\\\hline
\hline
\diagbox[width=2.6cm,height=1.1cm]{\raisebox{-1.5pt}{\hspace*{0.8cm}\Large$\boldsymbol{n}$}}{\raisebox{3.5pt}{\Large$\boldsymbol{z/d_{_{\rm Snell}}}$}\vspace{-0.3cm}\hspace*{-0.3cm}}&\hspace{0.35cm}\Large$\boldsymbol{100}$  &\hspace{0.4cm}\Large$\boldsymbol{500}$  & \hspace{0.25cm}\Large$\boldsymbol{1000}$ &\hspace{0.25cm}\Large $\boldsymbol{1500}$ &\hspace{0.16cm}\Large $\boldsymbol{2000}$ & \\\hline
\hline
\Large$\boldsymbol{\sqrt{2}}$ &\hspace{0.3cm}0.444& \hspace{0.3cm}0.472 &\hspace{0.25cm} 0.483 &\hspace{0.25cm} 0.475 & \hspace{0.3cm}0.460  &\vspace{-0.8cm}\hspace{0.4cm}\multirow{3}{*}{\rotatebox{-90}{\Large\textbf{ p\,-\,pol}}}\\
 \cline{1-6}
\Large $\boldsymbol{1.457} $& \hspace{0.3cm}0.440 &\hspace{0.2cm} 0.468 &\hspace{0.25cm} 0.480 & \hspace{0.35cm}0.472 & \hspace{0.3cm}0.457 &\\
 \cline{1-6}
\Large$\boldsymbol{1.515}$  &\hspace{0.2cm} 0.434 &\hspace{0.2cm} 0.463 & \hspace{0.35cm}0.475 & \hspace{0.35cm}0.469 & \hspace{0.3cm}0.454& \\\hline
\hline
&\multicolumn{5}{c||}{ \Large\bf $\bullet\,\,\,\,\boldsymbol{k\,\mbox{w}_{_0}=10^{^4}}\,\,\,\,\bullet\,\,\,\,\boldsymbol{\theta_{0}=\theta_{c}}\,\,\,\,\bullet\,\,\,\,\boldsymbol{N=50}\,\,\,\,\bullet$ }&\\\hline
    \end{tabular}
    }
\end{table}
\end{center}

\newpage

\begin{figure}[ht]
	\vspace{-1.99cm}
		\hspace{-0.8cm}
		\includegraphics[height=23.3cm, width=17cm]{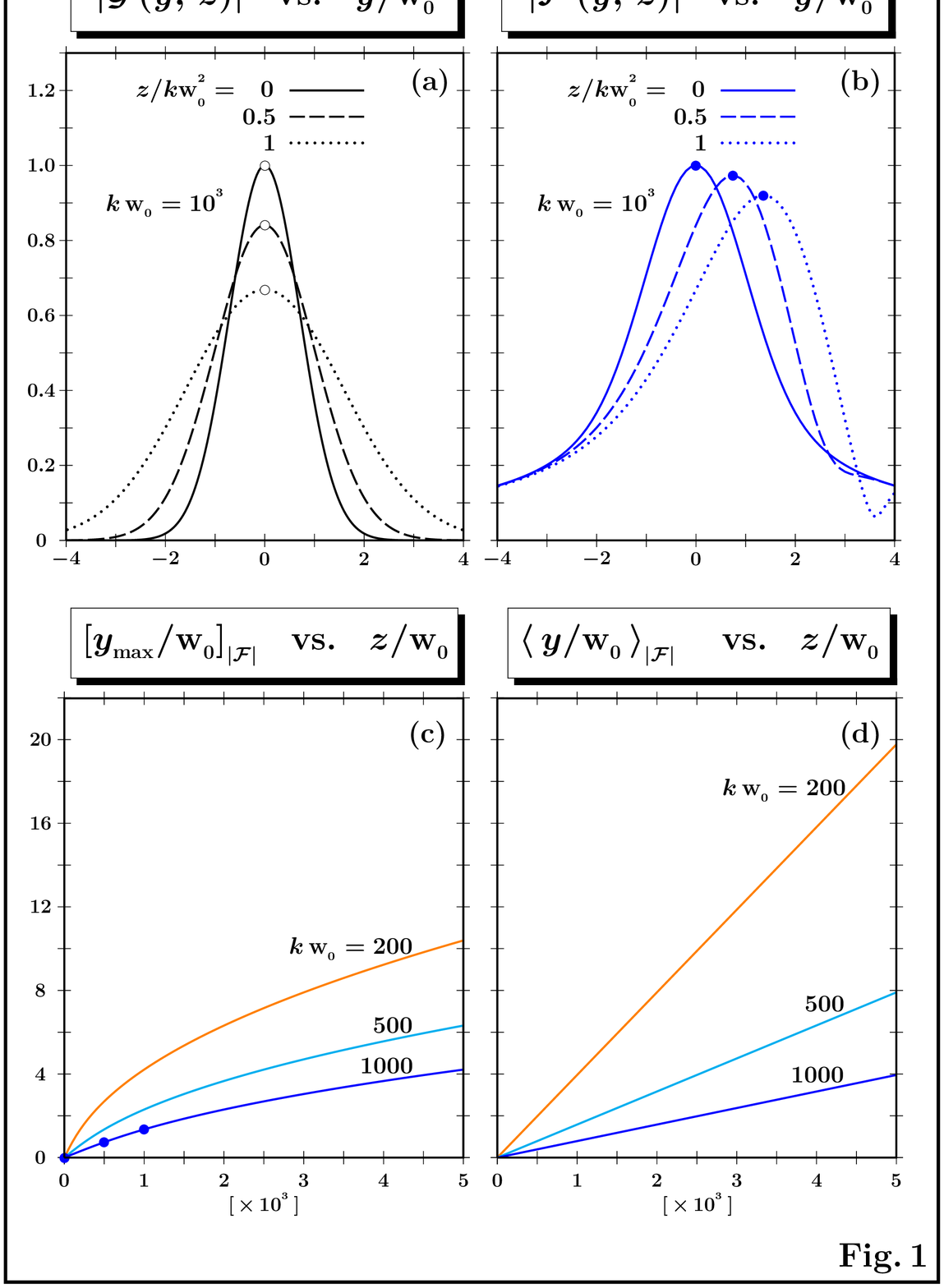}
		\vspace{-1cm}
\caption{\textbf{Modeled breaking of symmetry.} The breaking of symmetry in the gaussian angular distribution generates an axial dependence for the peak of the optical beam (b). This dependence is shown in (c). For
the transversal mean value is possible to obtain an axial linear analytical expression, given in  Eq.\,(\ref{dev}), which is confirmed by the numerical data plotted in (d).}
\label{fig:NSnellFig1}
\end{figure}

\newpage

\begin{figure}[ht]
	\vspace{-1.99cm}
		\hspace{-0.8cm}
		\includegraphics[height=23.3cm, width=17cm]{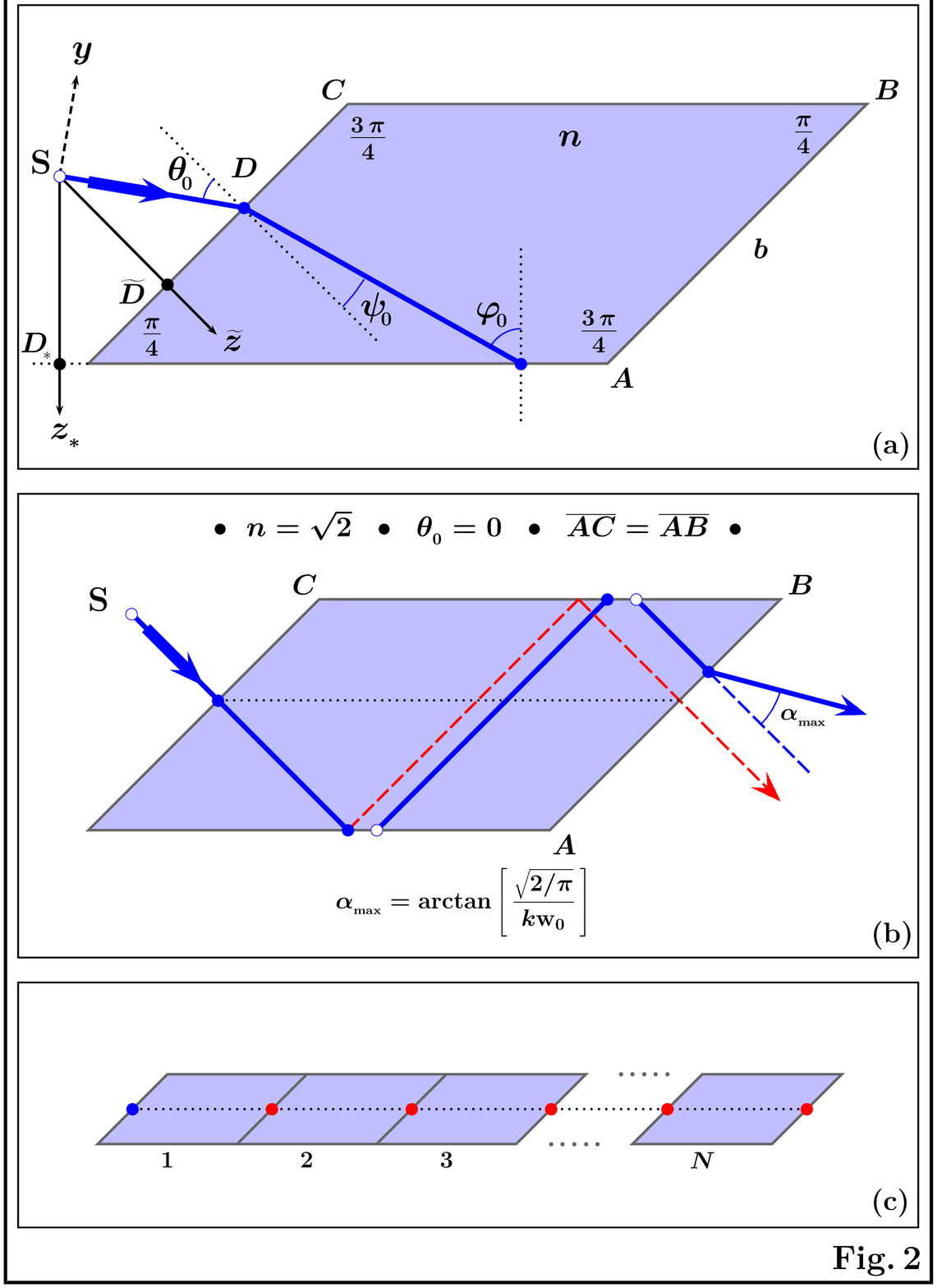}
		\vspace{-.8cm}
\caption{\textbf{Geometry of the dielectric block.} The normals to the left/right and up/down interfaces and the   angular parameters which appear in  the transmission coefficient are  given in (a). For a symmetric angular distribution the outgoing beam is parallel to the incoming one. The breaking of symmetry generates an angular deviation $\alpha$ of the Snell law which is drawn in (b) together the transversal Goos-H\"anchen shift.
The breaking of symmetry  is maximized by building a dielectric structure of $N$ blocks (c) which in a real optical experiment can be realized by a single elongated prism  of sides $N\, \overline{BC}$ and $\overline{AB}$.}
\label{fig:NSnellFig2}
\end{figure}

\newpage

\begin{figure}[ht]
	\vspace{-1.9cm}
		\hspace{-0.8cm}
		\includegraphics[height=23.3cm, width=17cm]{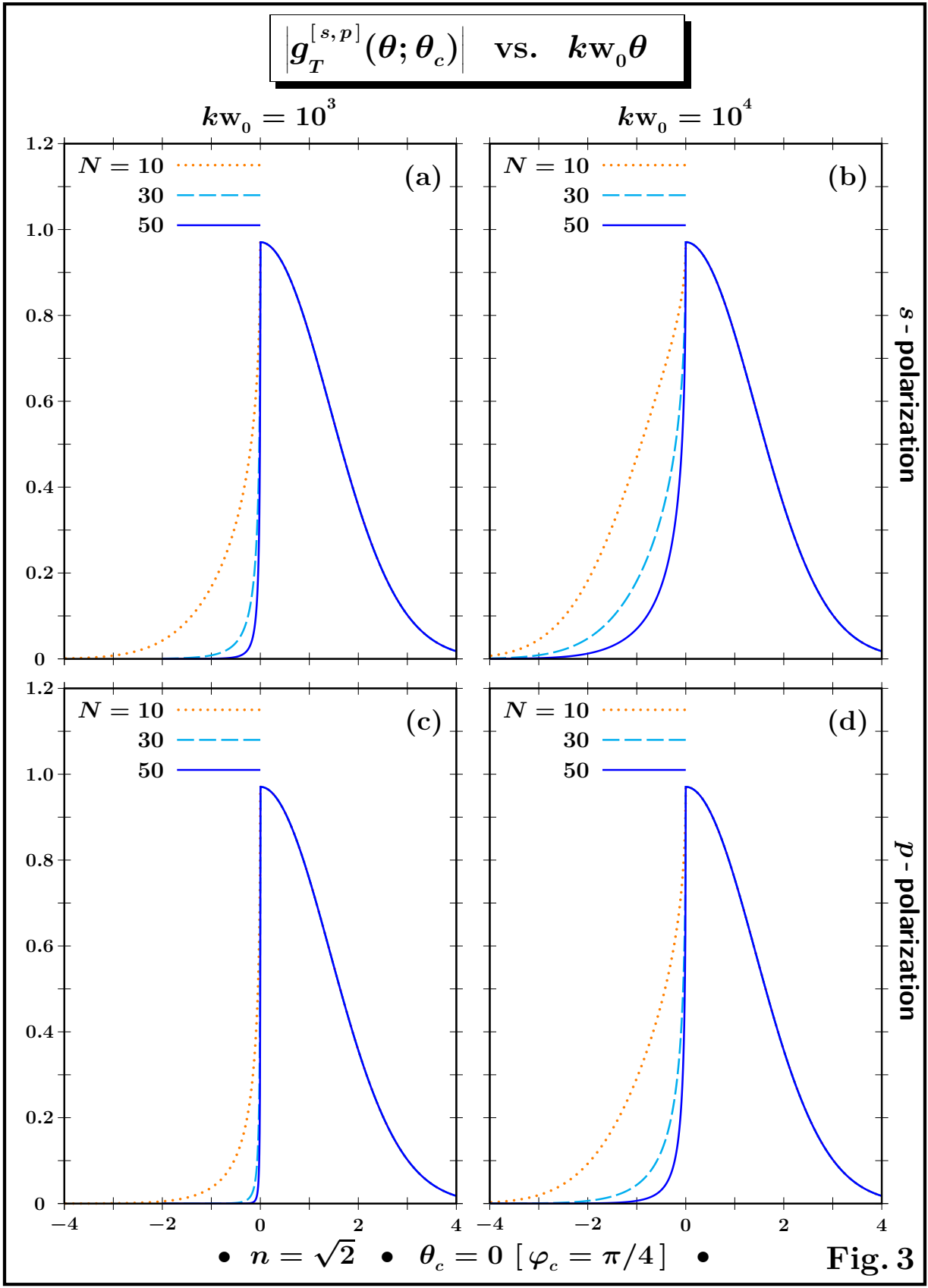}
		\vspace{-1cm}
		\caption{\textbf{Symmetry breaking for $\boldsymbol N$ dielectric blocks.} The modeled breaking of symmetry discussed in section II is now proposed for real optical experiments.  The plots  show that, to maximize the breaking of symmetry, we have to decrease the beam waist, increase the blocks number and use $p$-polarized waves. For p-polarized waves, an optimal choice to obtain a {\em maximal} breaking of symmetry is represented by $N=50$ and   $k{\rm w}_{\0}=10^{^{3}}$. To reproduce the maximal symmetry breaking for the other cases, we have to increase the number of blocks.}
\label{fig:NSnellFig4}
\end{figure}

\newpage

\begin{figure}[ht]
	\vspace{-1.9cm}
		\hspace{-0.8cm}
		\includegraphics[height=23.3cm, width=17cm]{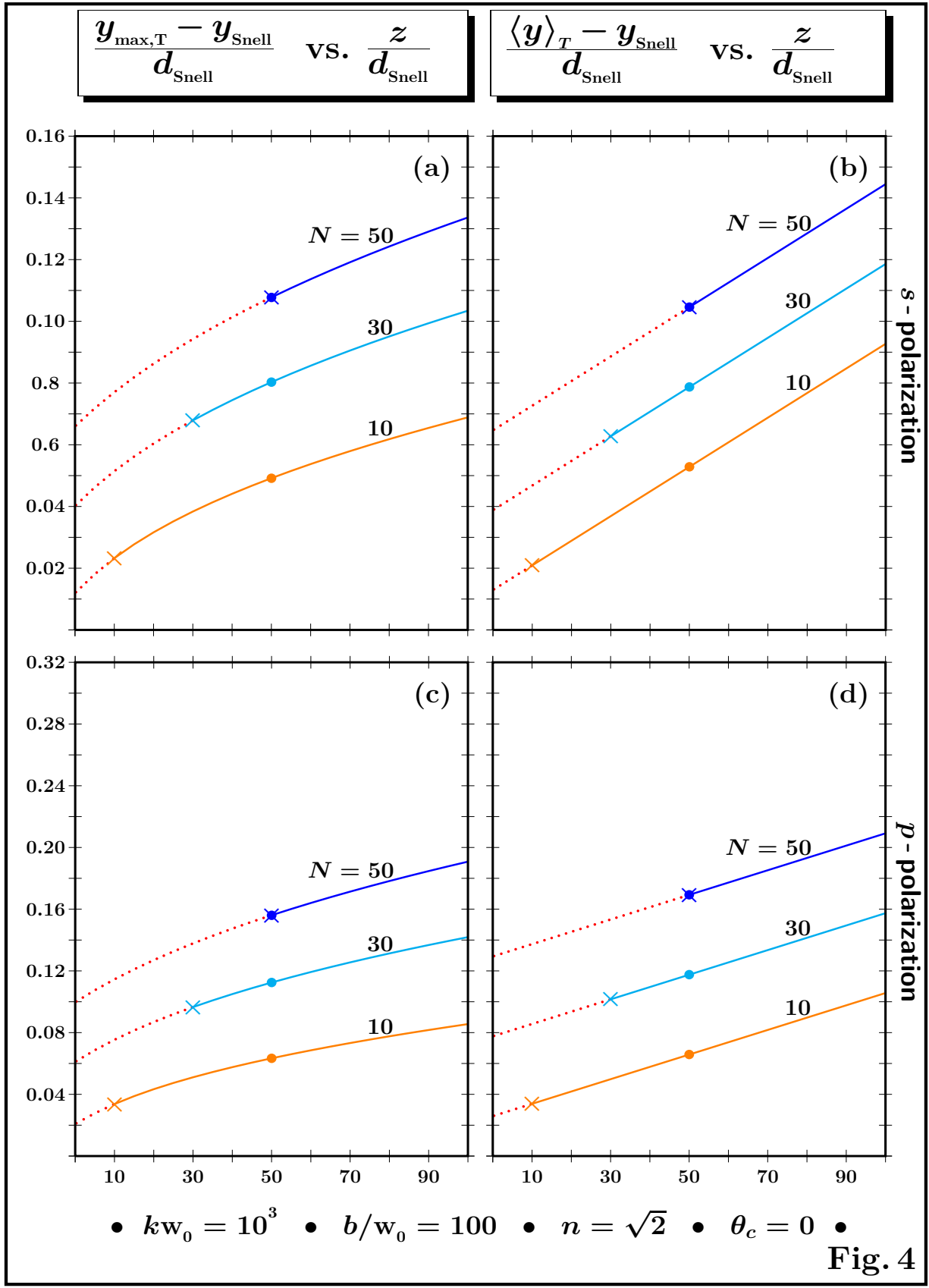}
		\vspace{-1cm}
		\caption{\textbf{Snell law angular deviation.} The axial dependence at critical angle of the peak and transversal mean value are plotted for a fixed beam waist w$_{\0}$ for different blocks number, $N$. The angular deviation of the Snell law is evident in (b) and (d). Observe that the first physical axial points
at which we can do the experimental analysis are given by  $z_{_{\rm out}} = z_{_{\rm in}} + N \tan \varphi_{c}\,\overline{AB}$\, ($\times$ points in the plots). From the plots is also clear the $N$-amplification of the Goos-H\"anchen shift at critical angle. The $\bullet$ points represent the points at which the numerical calculation has also been done for BK7 and Fused Silica block (Table 1). }
\label{fig:NSnellFig5}
\end{figure}

\newpage

\begin{figure}[ht]
	\vspace{-1.9cm}
		\hspace{-0.8cm}
		\includegraphics[height=23.3cm, width=17cm]{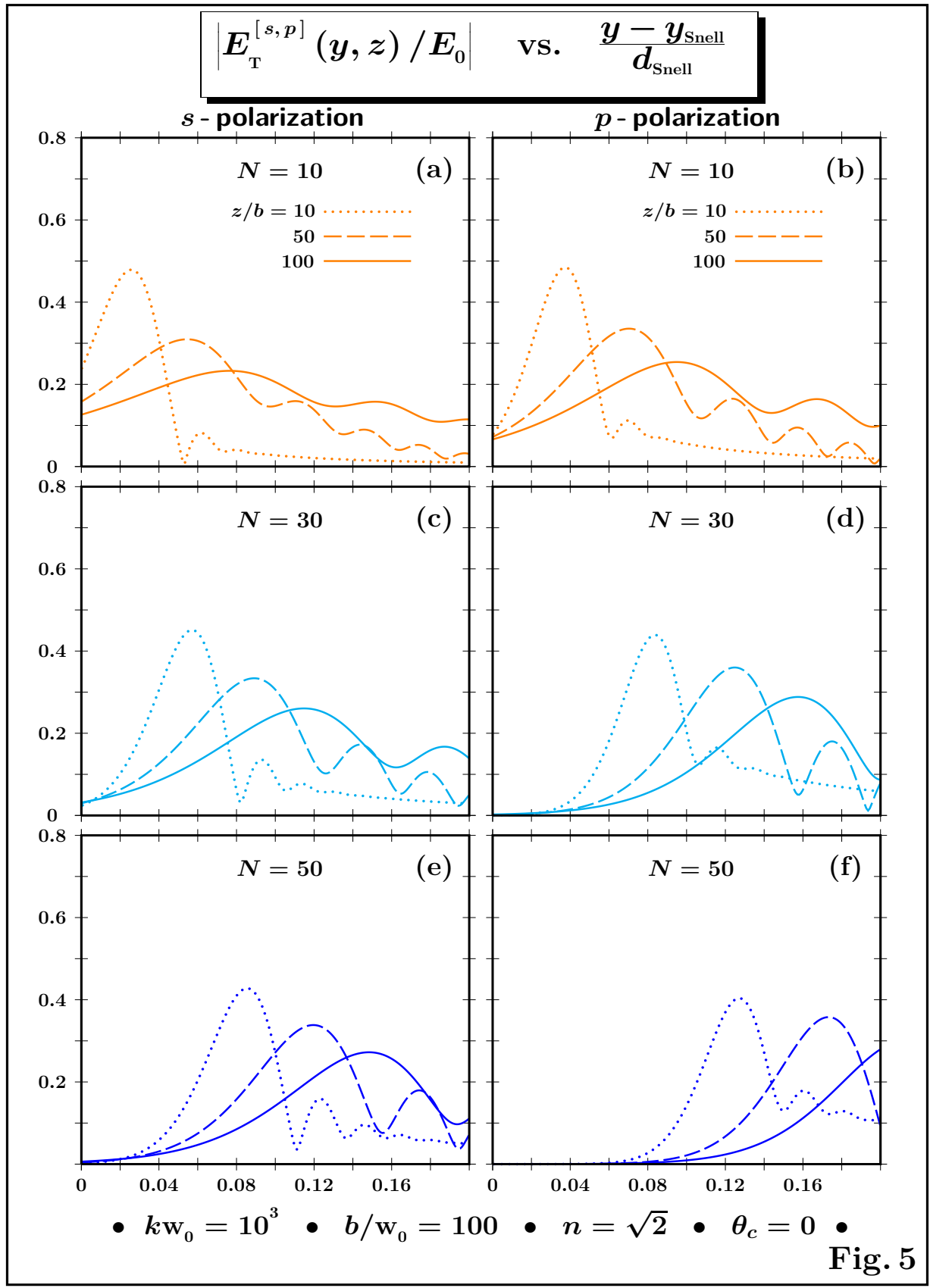}
		\vspace{-1cm}
		\caption{\textbf{Multiple peaks phenomenon.} For $k{\rm w}_{\0}=10^{^{3}}$, the outgoing optical beam presents the fascinating phenomenon of multiple peaks. This phenomenon is directly related to the spreading of the optical beam and it is due to the fact that in the angular distribution the positive angular components are no longer compensated by the negative ones.}
\label{fig:NSnellFig6}
\end{figure}

\newpage

\begin{figure}[ht]
	\vspace{-1.9cm}
		\hspace{-0.8cm}
		\includegraphics[height=23.3cm, width=17cm]{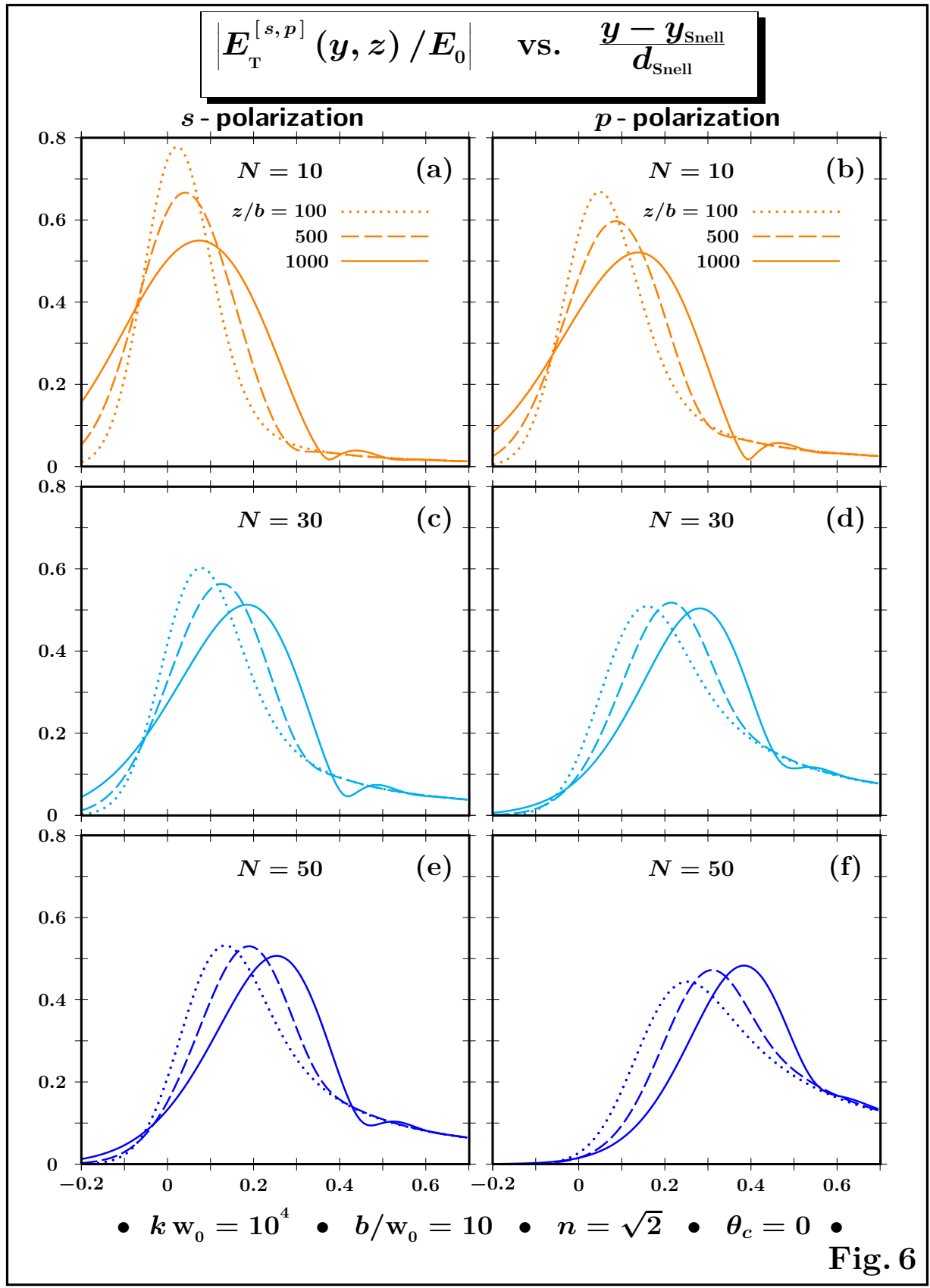}
		\vspace*{-1cm}
		\caption{\textbf{Focal effect.}  For $k{\rm w}_{\0}=10^{^{4}}$, the multiple peaks phenomenon is no longer so evident. Nevertheless, a new interesting phenomenon appears. Due to the second order optical phase contribution, in the outgoing beam can be now observed a focalization effect. This effect is for example clear as in (f).}
\label{fig:NSnellFig7}
\end{figure}

\end{document}